\documentclass[twocolumn]{aastex62}
\usepackage{epstopdf}
\usepackage{amssymb}
\received{\today}
\revised{\today}
\accepted{\today}
\submitjournal{ApJ}

\shorttitle{Mass-draining and Longitudinal Oscillations}
\shortauthors{Dai et al.}

\begin{document}
\title{Oscillations and mass-draining that lead to a sympathetic eruption of a quiescent filament}

\correspondingauthor{Jun Dai}
\email{daijun@pmo.ac.cn}

\author[0000-0002-0786-7307]{Jun Dai}
\affil{Key Laboratory of Dark Matter and Space Astronomy, Purple Mountain Observatory, CAS, Nanjing, 210023, People's Republic of China}
\affil{School of Astronomy and Space Science, University of Science and Technology of China, Hefei, 230026, People's Republic of China}

\author[0000-0003-4078-2265]{Qingmin Zhang}
\affiliation{Key Laboratory of Dark Matter and Space Astronomy, Purple Mountain Observatory, CAS, Nanjing, 210023, People's Republic of China}
\affiliation{School of Astronomy and Space Science, University of Science and Technology of China, Hefei, 230026, People's Republic of China}

\author{Yanjie Zhang}
\affiliation{Key Laboratory of Dark Matter and Space Astronomy, Purple Mountain Observatory, CAS, Nanjing, 210023, People's Republic of China}
\affiliation{School of Astronomy and Space Science, University of Science and Technology of China, Hefei, 230026, People's Republic of China}

\author[0000-0002-9121-9686]{Zhe Xu}
\affiliation{Key Laboratory of Dark Matter and Space Astronomy, Purple Mountain Observatory, CAS, Nanjing, 210023, People's Republic of China}
\affiliation{School of Astronomy and Space Science, University of Science and Technology of China, Hefei, 230026, People's Republic of China}

\author{Yingna Su}
\affiliation{Key Laboratory of Dark Matter and Space Astronomy, Purple Mountain Observatory, CAS, Nanjing, 210023, People's Republic of China}
\affiliation{School of Astronomy and Space Science, University of Science and Technology of China, Hefei, 230026, People's Republic of China}

\author{Haisheng Ji}
\affiliation{Key Laboratory of Dark Matter and Space Astronomy, Purple Mountain Observatory, CAS, Nanjing, 210023, People's Republic of China}
\affiliation{School of Astronomy and Space Science, University of Science and Technology of China, Hefei, 230026, People's Republic of China}

\begin{abstract}

In this paper, we present a multi-wavelength analysis to mass-draining and oscillations in a large quiescent filament prior to its successful eruption on 2015 April 28.
The eruption of a smaller filament that was parallel and in close,
$\sim$350$\arcsec$ proximity was observed to induce longitudinal oscillations and enhance mass-draining within the filament of interest.
The longitudinal oscillation with an amplitude of $\sim$25 Mm and $\sim$23 km s$^{-1}$
underwent no damping during its observable cycle.
Subsequently the slightly enhanced draining may have excited a eruption behind the limb,
leading to a feedback that further enhanced the draining and induced simultaneous oscillations within the filament of interest.
We find significant damping for these simultaneous oscillations,
where the transverse oscillations proceeded with the amplitudes of $\sim$15 Mm and $\sim$14 km s$^{-1}$,
while the longitudinal oscillations involved a larger displacement and velocity amplitude ($\sim$57 Mm, $\sim$43 km s$^{-1}$).
The second grouping of oscillations lasted for $\sim$2 cycles and had the similar period of $\sim$2 hours.
From this, the curvature radius and transverse magnetic field strength of the magnetic dips
supporting the filaments can be estimated to be $\sim$355 Mm and $\geq$34 G.
The mass-draining within the filament of interest lasted for $\sim$14 hours.
The apparent velocity grew from $\sim$35 km s$^{-1}$ to $\sim$85 km s$^{-1}$,
with the transition being coincident with the occurrence of the oscillations.
We conclude that two filament eruptions are sympathetic,
i.e. the eruption of the quiescent filament was triggered by
the eruption of the nearby smaller filament.

\end{abstract}


\section{Introduction} \label{sec:intro}
Solar filaments are features filled with cool and dense plasma suspended in the corona \citep{lab10,mac10,reg11}.
They usually form along the magnetic polarity inversion lines (PILs) and can be erupted \citep{vanb89,mat98,gib18}.
Frequently, two adjoining filaments could erupt successively.
This kind of phenomenon is termed as a sympathetic eruption.
The sympathetic eruptions appear to occur sequentially over relatively short periods of time,
across separated source regions \citep{liu2009}, and sometimes even across a full hemisphere \citep{zhukov2007}.
They are usually interlinked by large-scale background magnetic fields \citep{cheng2005,Schrijver2011,Titov2012,Schrijver2013},
and physical connections of magnetic nature between sympathetic eruptions have been confirmed by
statistical analysis \citep{moon2002,wheatland2006} and case studies \citep{wang2001,ding2006,wang2016}.
Generally, coronal mass ejections (CMEs), extreme ultraviolet (EUV) waves, and the propagating perturbations along magnetic field can cause
sympathetic eruptions \citep{wang2001,jiang2008,tor2011,lynch2013,jin2016}.
In certain case of sympathetic eruptions between multiple filaments,
the physical connection is interpreted as the magnetic reconnection between the overlying magnetic field \citep{tor2011,shen2012,wang2018,song2020,hou2020}.

Apparently, single filament eruption is more common than sympathetic eruptions.
Since the filaments are highly dynamic, the destabilization of a filament may lead to an eruption
as a result of magnetic reconnection or ideal magnetohydrodynamic (MHD) instabilities \citep[]{for06,chen11,jan15}.
Moreover, the total mass of a filament reaches up to 10$^{14}$--10$^{15}$ g \citep{par14},
and the mass of a filament is believed to be important for its stabilization  \citep{zhang2021arXiv}.
\citet{Low1996SoPh} proposed that the decrease in weight could cause a filament to rise up and finally erupt.
Using simultaneous observations in H$\alpha$ and He {\sc i} 10830 {\AA},
\citet{Gilbert2001ApJ} observed fragmentary disappearance of a filament as the plasma flows onto the disk.
The stereoscopic observation on 2010 April 3 shows that the slow uplift and eruption of the filament are induced by mass-unloading \citep{sea11}.
Recently, \citet{jenk18} also presented an analysis for stereoscopic observation of the partial eruption of a quiescent prominence
and concluded that impulsive mass-unloading is responsible for the eruption.
Moreover, \citet{jenk19} modeled a simple flux rope to quantify the effect of mass-draining in a filament.
It is revealed that the rapid depletion of mass before the loss of equilibrium facilitates the increase in height of the flux rope.
\citet{fan20} performed 3D MHD numerical simulations for mass-unloading in a filament, they found that the presence of mass within the flux rope lead to an increase in the height required for loss-of-equilibrium and mass-draining can cause an earlier eruption by $\sim$6 hr.
These observations and numerical simulations suggest that mass-draining may play an important role in the filament eruption.

Apart from mass-draining, a filament will exhibit various kinds of oscillatory motions
when it is being disturbed \citep{ball06,Oliver2009,Tripathi2009SSRv,arr18}.
According to the velocity amplitude, filament oscillations are generally classified as small-amplitude ($\leq$10 km s$^{-1}$)
and large-amplitude ($\geq$20 km s$^{-1}$) oscillations \citep{Oliver2002,luna18}.
Depending on their directions,
the large-amplitude oscillations (LAOs) are further divided into
large-amplitude transverse oscillations (LATOs) perpendicular to the axis of a corresponding filament
and large-amplitude longitudinal oscillations (LALOs) along the axis.
Generally, LATOs are triggered by global magnetohydrodynamic (MHD) waves or shock waves,
 such as chromospheric Moreton waves \citep{Moreton1960} and coronal EUV waves \citep{Okamoto2004},
 which are closely associated with remote flares or CMEs \citep{Hyder1966,Ramsey1966,bocc11,liu2013,shen14a,shen17,zhang2018}.
In a particular case, slow rise or pre-eruption of a nearby erupting filament could also excite LATOs \citep{iso06,Isobe2007,pinter2008,chen08}.
The magnetic tension is theoretically considered as the main restoring force for LATOs,
and the damping is usually due to the energy loss or dissipative process \citep{Kleczek1969SoPh,Tripathi2009SSRv}.
LALOs are mostly triggered by solar flares adjacent to the filaments footpoints
\citep{jing03,jing06,vrs07,LZ12,zhang12,zhang17b,zhang20} or by coronal jets \citep{luna14,zhang17a,luna21}.
Sometimes, LALOs could be triggered by coronal shock waves during flares \citep{shen14b,Pant2015RAA}
or by the merging of two adjacent filaments \citep{luna2017}.
The predominant restoring force of LALOs is believed to be the projected gravity along the flux tube supporting the filament threads \citep{Luna2012,zhang12,zhou17}.
The primary damping mechanisms for LALOs include mass accretion \citep{LK12,rud16,awa19},
radiative loss \citep{zhang13}, and wave leakage \citep{zhang19}.
A simulation shows that mass drainage reduces the damping time considerably in one strong perturbation \citep{zhang13}.
Recent numerical simulations have shed light on the nature of large-amplitude longitudinal oscillations \citep[e.g.,][]{ter15,zhou18,adro20,li20,Liakh2021}.

LAOs before eruptions are frequently observed,
showing that filament oscillation can serve as a kind of precursor for subsequent eruption \citep{chen08,zhang12}.
\citet{bi14} presented early reports of both LALOs and mass-draining preceding a filament eruption,
and the similar processes have been nicely reproduced with 3D numerical simulations \citep{fan20}.
However, questions like whether there is a mutual facilitation between the filament oscillation and mass-draining,
and whether both contribute to the triggering a filament eruption are still unclear and need more analysis with well-observed cases.

On 2015 April 28, a sympathetic successful filament eruption induced by a nearby filament eruption occurred in the northeast hemisphere of the Sun.
For the same event, \citet{lor21} studied the plasma outflows originating in the coronal dimming after the sympathetic eruption.
In this paper, we focus on the oscillations and the mass-draining before the successful eruption.
The observation and data analysis are described in Section~\ref{sec:obs}.
The results are presented in Section~\ref{sec:res}.
A comparison with previous findings and a brief conclusion are given in Section~\ref{sec:dis}.

\begin{figure*}
\centering\includegraphics[width=0.855\textwidth]{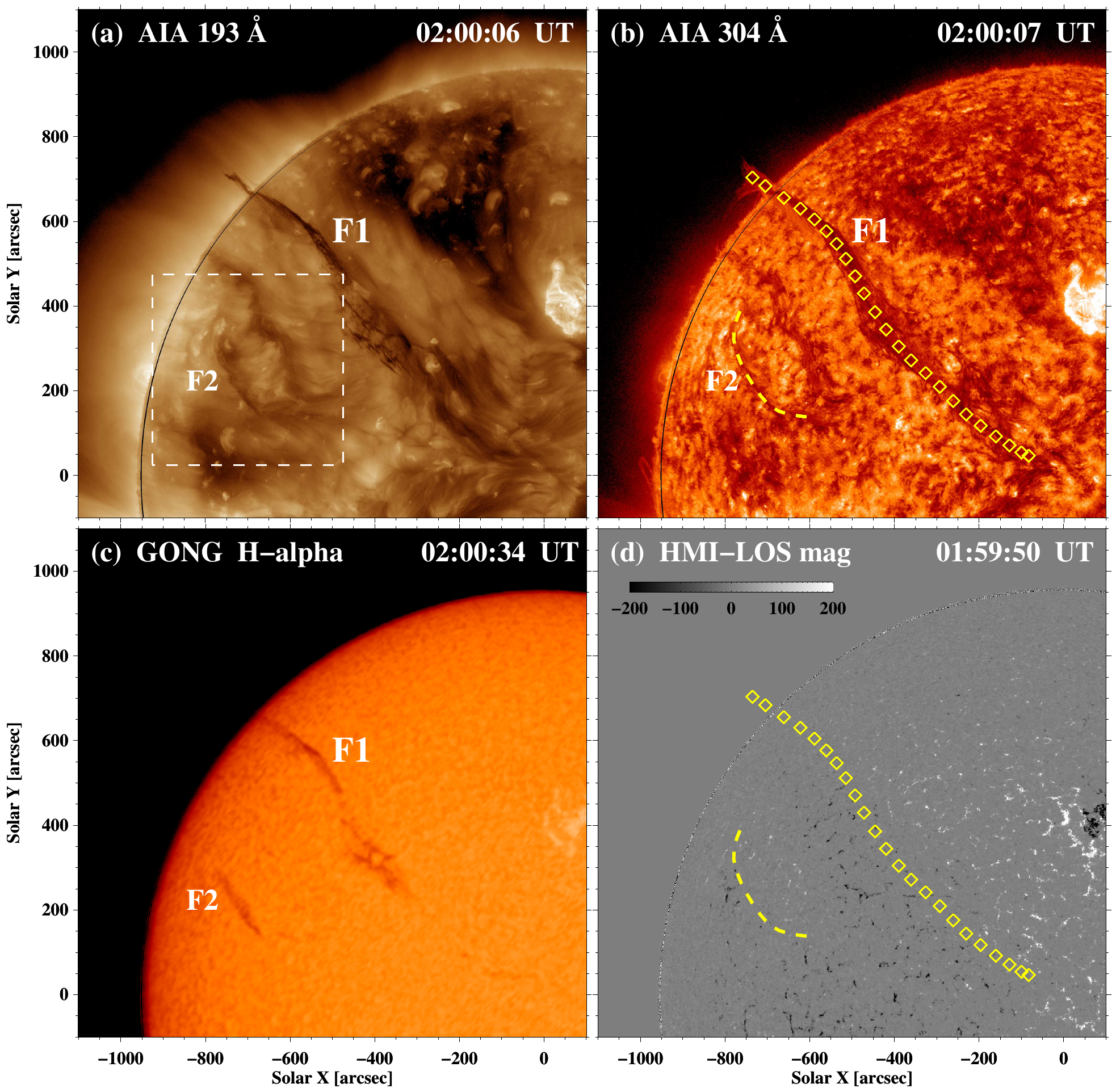}
\caption{\small{The appearance of the two filaments labeled as F1 and F2 in AIA 193{~\AA} (a), 304{~\AA} (b) and H$\alpha$ (c) images.
               Yellow dashed line outlines F2 and the yellow diamonds outline F1.
               Panel (d) give an HMI LOS magnetogram with the magnetic field strengths in the range of -200 and 200 G.
                The white rectangle in panel (a) gives the field of view (FOV) of Figure~\ref{fig2}.
                 }}
\label{fig1}
\end{figure*}

\section{Observation and data analysis} \label{sec:obs}
On 2015 April 28, two quiescent filaments, located in the northeast quadrant of the solar disk,
were simultaneously observed by the Global Oscillation Network Group (GONG)
in H$\alpha$ line center (6562.8{~\AA})
and by the Atmospheric Imaging Assembly \citep[AIA;][]{lemen12}
on board the Solar Dynamics Observatory \citep[SDO;][]{pesn12}.
The full-disk H$\alpha$ images have a cadence of 60 s and a spatial resolution of 2$\arcsec$.
AIA takes full-disk images in two ultraviolet (UV; 1600 and 1700{~\AA}) wavelengths with a cadence of 24 s and in seven EUV (94, 131, 171, 193, 211, 304, and 335{~\AA})
wavelengths with a cadence of 12 s.
The photospheric line-of-sight (LOS) magnetograms were observed by the Helioseismic
and Magnetic Imager \citep[HMI;][]{schou12} on board SDO with a cadence of 45 s.
The level\_1 data from AIA and HMI with a spatial resolution of 1$\farcs$2 were calibrated using the standard Solar SoftWare (SSW) programs \texttt{aia\_prep.pro} and \texttt{hmi\_prep.pro}.
The images observed in H$\alpha$ and 304{~\AA} were co-aligned with the cross-correlation method using sunspots as references.
The large-scale 3D magnetic configuration near the filaments was derived from the potential field source surface \citep[PFSS;][]{Schrijver2003SoPh} modeling.
The associated CMEs were observed by the C2 on board the SOHO Large Angle and Spectrometric Coronagraph \citep[LASCO;][]{bru95}.

\begin{figure*}[htb]
\centering\includegraphics[width=0.88\textwidth]{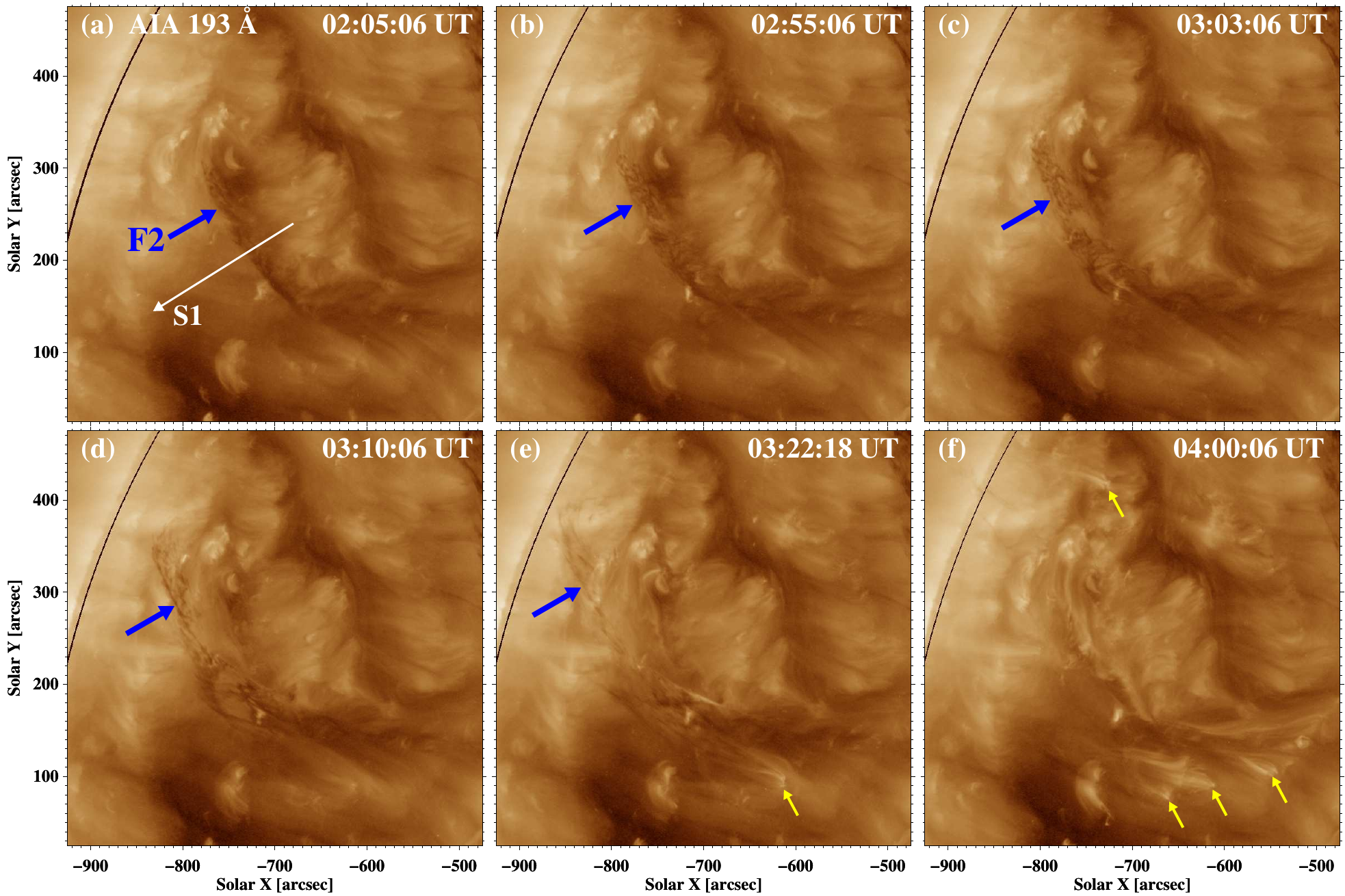}
\caption{Six snapshots of the AIA 193{~\AA} images, showing the temporal evolution of the eruption of F2.
The blue arrows point to the spine of F2, and the yellow arrows point to the brightenings at the footpoints of F2.
In panel (a), the white slice S1 is used to investigate the height evolution of F2.
An animation (\textit{F2eruption.mp4}) of the unannotated SDO observations is available.
\textit{F2eruption.mp4} covers $\sim$4 hr starting at 00:59:54 UT and ending the same day at 04:55:30 UT, with time cadence of 24 seconds.
The FOV is displayed by the white rectangle in Figure~\ref{fig1}(a)
(An animation of this figure is available.)
\label{fig2}}
\end{figure*}

\section{Results} \label{sec:res}

\subsection{F2 eruption and another eruption} \label{sec:fail}

In Figure~\ref{fig1}, the top panels show EUV images in 193{~\AA}
and 304{~\AA} at $\sim$02:00:06 UT.
The bottom left panel shows the H$\alpha$ image at 02:00:34 UT with the same field of view (FOV).
Two parallel quiescent filaments (F1 and F2) are clearly seen and F1 is much longer than F2.
The distance between the two filaments is $\sim$350$\arcsec$.
Figure~\ref{fig1}(d) shows the photospheric LOS magnetogram at 01:59:50 UT,
with the spines of F1 and F2 being marked with a series of yellow diamonds and a dashed line, respectively.
It is clear that both filaments are located along PILs.

From 02:00 UT to 15:00 UT, F2 and F1 erupted successively,
which could be considered as two separate eruptions due to the long time span.
In addition, there was another filament eruption behind F1 during 06:00$-$08:00 UT (see also the online animation \textit{F1eruption.mp4}),
which is labeled as Eruption3 in Figure~\ref{fig4}(a).
Figure~\ref{fig2} shows six snapshots of EUV images in 193{~\AA} during 02:00$-$04:00 UT,
where the spine of F2 is indicated by the blue arrows (see also the online animation \textit{F2eruption.mp4}).
It shows that F2 rose slowly from $\sim$01:05 UT and reached its eruption apex at $\sim$03:20 UT.
The erupted material appears to have fallen back towards the two footpoints and caused brightenings,
as indicated by the yellow arrows in Figure~\ref{fig2}(e-f).
Hence, F2 may be defined as a failed eruption \citep{ji03,dai21}.
However, the presence of a weak associated CME can be identified within the LASCO/C2 snapshots of Figure~\ref{fig7} (a1-a3),
indicating this was in fact a successful eruption.

\begin{table*}
\centering
\caption{Fitted parameters for longitudinal and transverse filament oscillations.}
\label{tab-1}
\begin{tabular}{c|cccccccc}
  \hline
    & $t_{sta}$ & $t_{end}$  &$A_0$  & $P$  & $\tau$ & $\tau/P$ & $v_{max}$\\
    &     (UT)      & (UT)&    (Mm)     &(min)   &(min) &      & (km s$^{-1}$) \\
     \hline
  OS1 & 03:35 & 07:10       & 25.1  & 114  &  ... &   ...   &     23    \\
   \hline
  OS2 & 06:05 & 10:25       & 57.5  & 142  & 230     &1.6  &   46  \\
   \hline
  OS3 & 07:20 & 10:35       & 56.9  & 126  & 168     &1.3  &   41 \\
   \hline
  OS4 & 06:05 & 09:55        & 14.7  & 105  & 142     &1.4  &   14 \\
   \hline
\end{tabular}
\end{table*}

To investigate the height evolution of F2,
we select a straight slice (S1) with a length of 179$\arcsec$ along the eruption direction in Figure~\ref{fig2}(a).
The time-space diagram of S1 in 193{~\AA} is plotted in Figure~\ref{fig5}(a)
and the height of F2 can be obtained by outlining the edge in the the diagram.
The height evolution in the plane-of-sky is characterized by a slow rise with a constant speed followed by an initial impulsive acceleration \textbf{indicated} by an apparent exponential increase. Hence, we fit $h(t)$ using the function as proposed by \cite{cheng13}:
\begin{equation} \label{eqn-1}
  h(t) = c_{0}e^{(t-t_{0})/{\tau}} + c_{1}(t-t_{0}) + c_{2},
\end{equation}
where $t_0$, $\tau$, $c_0$, $c_1$, and $c_2$ are free parameters.
This model is composed of of a linear term and an exponential term,
which correspond to the slow rise and the impulsive acceleration phase, respectively.
The exponential term is reasonable, since it describes the impulsive acceleration of the filament
when it is triggered by the flare reconnection \citep[e.g.,][]{Moore2001}
or the MHD instability \citep[e.g.,][]{tor2005}.
The onset of fast-rise phase is defined by setting the parameters as the point when the exponential velocity is equal to the linear velocity:
\begin{equation} \label{eqn-2}
  t_{onset} = \tau\ln(c_1 \tau/c_0)+t_0.
\end{equation}
The fitting process is implemented by the standard SSW program \texttt{mpfit.pro} and the fitted curve is plotted with the yellow solid line in Figure~\ref{fig5}(a).
It is clear that the fitted curve can nicely represent the height evolution of F2.
The spine of F2 started to rise slowly from $\sim$01:05 UT at a speed of $\sim$2 km s$^{-1}$.
The fast rise began at $\sim$02:30 UT and lasted for $\sim$50 minutes with an average speed of $\sim$85 km s$^{-1}$.

\begin{figure}
\centering\includegraphics[width=0.47\textwidth]{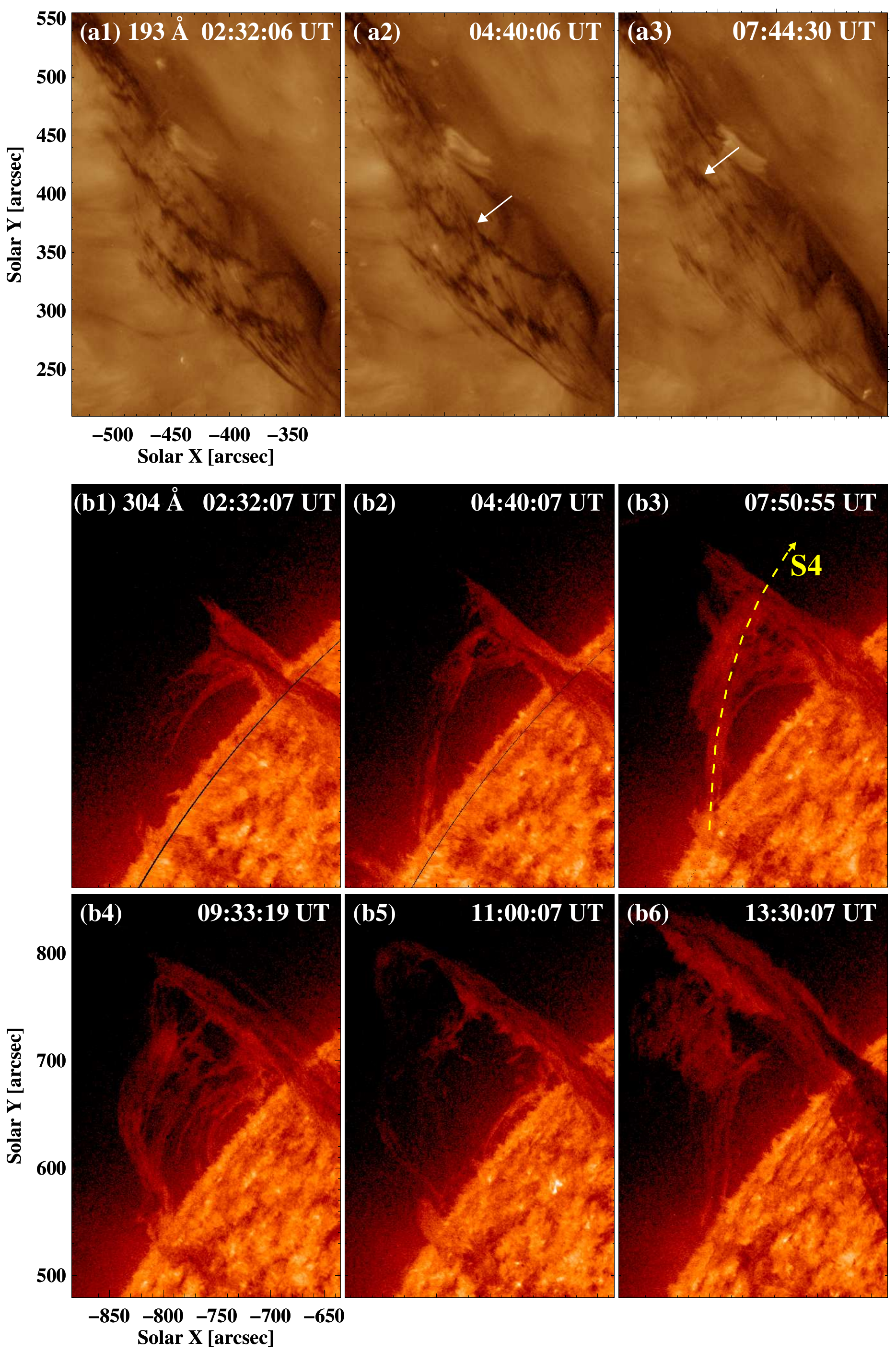}
%\plotone{figure3.eps}
\caption{Panels (a1-a3) give three AIA 193{~\AA} images, showing the longitudinal oscillations in F1.
The white arrows point to an oscillating thread.
Panel (b1-b6) give six snapshots of AIA 304{~\AA} images, showing the mass-draining toward the solar surface along the barb.
The curved slice S4 is used to investigate the mass-draining.
An animation (\textit{OSandDR.mp4}) of the unannotated SDO observations is available.
\textit{OSandDR.mp4} covers $\sim$13 hr starting at 01:00:18 UT and ending the same day at 13:59:55 UT, with time cadence of 1 minute.
The FOV is displayed by the white rectangles in Figure~\ref{fig4}(b).
(An animation of this figure is available.)
\label{fig3}}
\end{figure}

\subsection{Oscillations in F1} \label{sec:osc}

\begin{figure*}
\centering\includegraphics[width=0.825\textwidth]{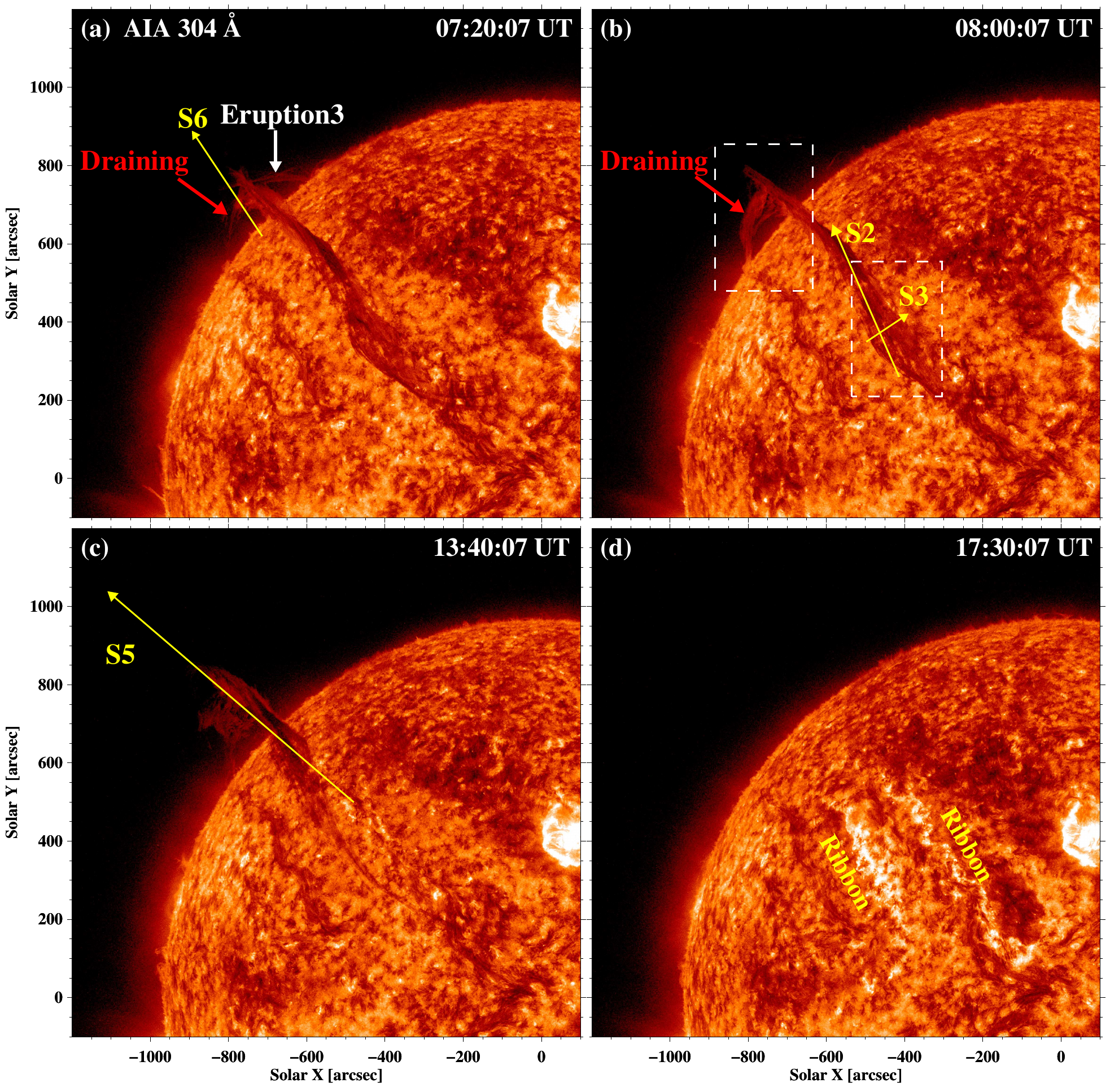}
\caption{Four snapshots of AIA 304{~\AA} images showing the mass-draining and eruption of the larger filament.
The red arrows point to the mass-draining along the barb.
The white arrow points to another eruption (Eruption3) near the barb,
which is traced by the yellow slice S6.
In panel (b), the dashed boxes denote the FOV of Figure~\ref{fig3}(a1-a3) and Figure~\ref{fig3}(b1-b6), respectively.
The yellow slice S2 and S3 are used to investigate the longitudinal and transverse oscillations in F1.
The yellow slice S5 in panel (c) is used to investigate the slow rise and fast rise of F1.
An animation (\textit{F1eruption.mp4}) of the unannotated SDO observations is available.
\textit{F1eruption.mp4} covers $\sim$16.7 hr starting at 00:00:07 UT and ending the same day at 16:40:31 UT, with time cadence of 1 minute.)
(An animation of this figure is available.)
\label{fig4}}
\end{figure*}

\begin{figure*}
\centering\includegraphics[width=0.825\textwidth]{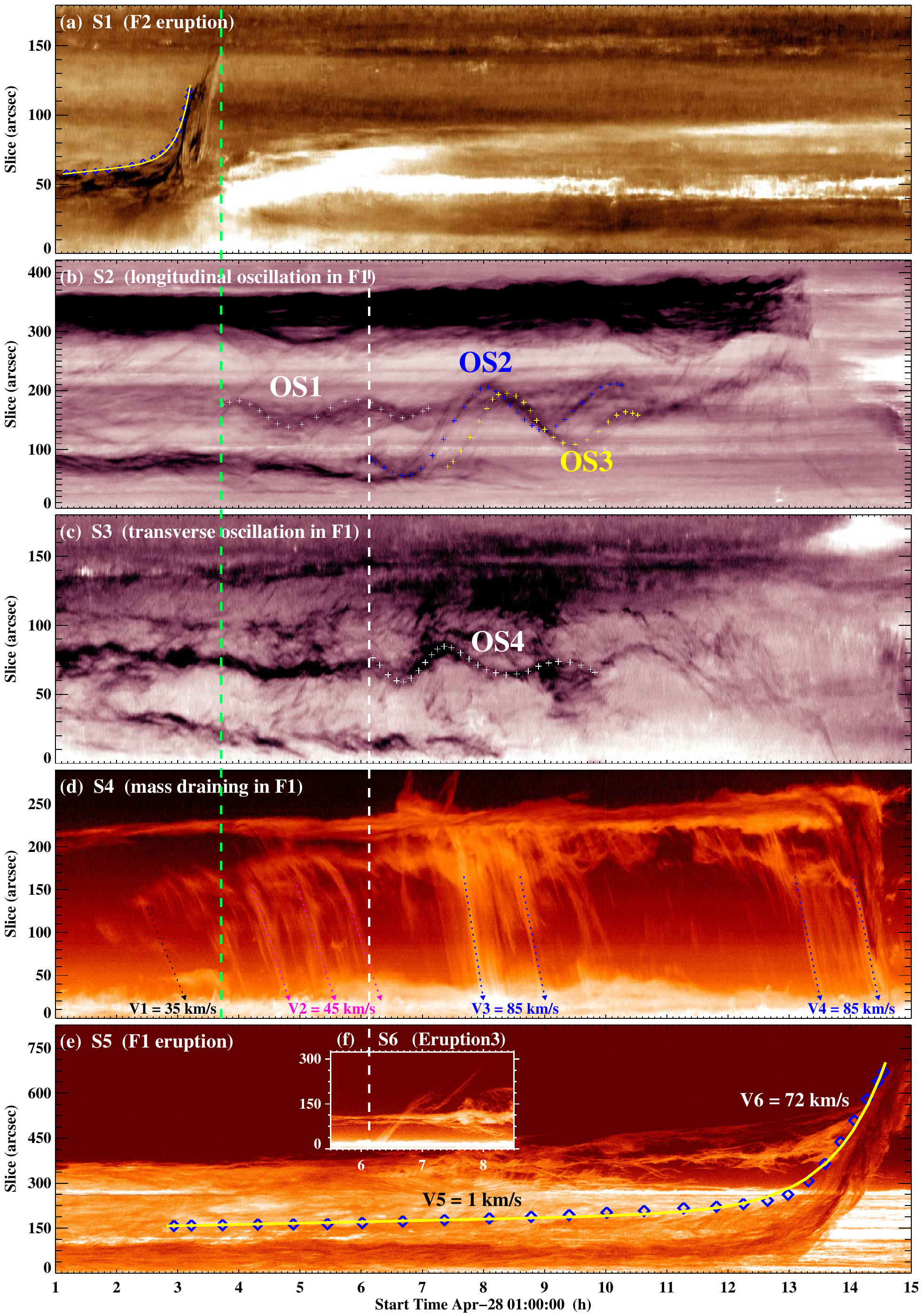}
\caption{Time-space diagrams for slice S1-S6 as given in previous figures.
The white, blue, and yellow dashed lines in panel (b) outline the longitudinal oscillations.
The white dashed line in panel (c) outlines the transverse oscillation.
The black, magenta, and blue dotted lines in panel (d) are used to calculate the apparent velocities of mass-draining at different phases.
The green dashed line in panels (a-d) marks the end time of F2 eruption.
The white dashed line in panels (b-f) marks the start time of the simultaneous oscillations (OS2,OS4) and Eruption3.
The blue diamonds in panel (a) and (e) outline the heights of F2 and F1,
and the white and yellow solid line represents the fitted curve using Equation~\ref{eqn-1}.
The apparent velocities during the slow rise and fast rise are labeled.
\label{fig5}}
\end{figure*}

After the cool material in F2 fell back to the footpoints at $\sim$03:35 UT, the threads in F1 began to oscillate along its axial direction,
which is indicated by white arrows in Figure~\ref{fig3}(a2-a3) (see also the online movie \textit{OSandDR.mp4}).
To investigate the longitudinal oscillations and whether there are transverse oscillations in F1,
we select a straight slice (S2) along the axis with a length of 421$\arcsec$
and another straight slice S3 perpendicular to the axis with a length of 180$\arcsec$ in Figure~\ref{fig4}(b),
and the two time-space diagrams in 211{~\AA} are plotted in Figure~\ref{fig5}(b-c).
Surprisingly, we find a set of transverse oscillating signals, which is obviously damping.
The transverse damping oscillation (OS4) during 06:05$-$09:55 UT is drawn with white pluses in Figure~\ref{fig5}(c).
In addition, we find three sets of longitudinal oscillating signals, two are damping while one has no obvious damping (non-damping).
The non-damping oscillation (OS1) lasts for $\sim$2 cycles during 03:35$-$07:10 UT is drawn with white pluses.
The damping oscillations (OS2 and OS3) are drawn with blue and yellow pluses, respectively.
The simultaneous oscillations (i.e. OS2 and OS4) start almost at the same time ($\sim$06:05 UT)
and both fade out after going through $\sim$2 cycles.
The OS3 lags behind OS2 by $\sim$75 minutes and last for $\sim$1.5 cycles.
To investigate the initiation of Eruption3,
we select a straight slice (S6) with a length of 324$\arcsec$ and
the time-space diagram in 304{~\AA} is plotted in Figure~\ref{fig5}(f).
In Figure~\ref{fig5}(b-f),
the white dashed line denotes the start time of OS2 and OS4 (06:05 UT) as well as Eruption3.
Therefore, OS1 started after F2 eruption
while the simultaneous oscillations OS2 and OS4 started after the onset of Eruption3.

To precisely determine the parameters of oscillations in F1,
we fit the curves in Figure~\ref{fig5}(b-c) using \texttt{mpfit.pro} and the following function:
\begin{equation} \label{eqn-3}
  A(t) = A_{0}\sin \bigg(\frac{2\pi t}{P} + \psi \bigg)e^{-\frac{t}{\tau}} + A_{1}t + A_{2},
\end{equation}
where $A_0$ is the initial amplitude, $P$ is the period, $\tau$ is the damping time, $\psi$ is the initial phase,
and $A_{1}t+A_{2}$ represents a linear term of the equilibrium position of the filament threads.

\begin{figure}
\centering\includegraphics[width=0.475\textwidth]{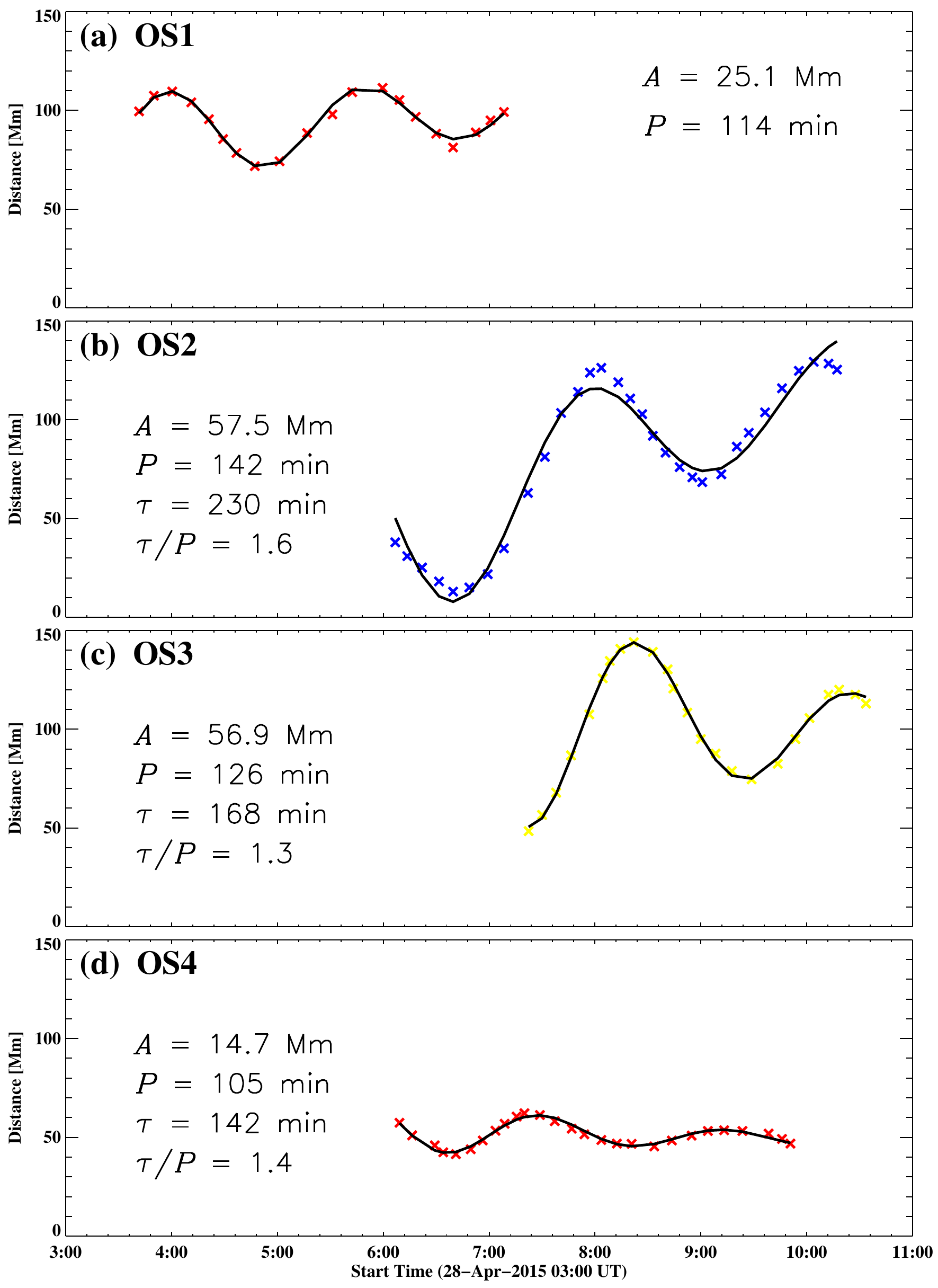}
\caption{Extracted positions of the oscillating threads during OS1-OS4.
The fitted curves are overlaid with black solid lines.
Corresponding parameters are labeled in each panel.
\label{fig6}}
\end{figure}

In Figure~\ref{fig6}, the four groups of crosses represent the extracted positions of the oscillating filament threads along S2 and S3 in 211 {~\AA}.
The results of curve fitting using Equation~\ref{eqn-3} are overlaid with black solid lines.
It is evident that the curve fitting is satisfactory, and the fitted parameters are listed in Figure~\ref{fig6} and Table~\ref{tab-1}.
For the longitudinal oscillations, OS1 has an amplitude of $\sim$25.1 Mm and a period of $\sim$114 min.
OS2 and OS3 have displacement amplitudes of 57$\pm$0.5 Mm, which is $\sim$2.3 times larger than that of OS1.
The velocity amplitudes (41$-$46 km s$^{-1}$) are $\sim$2 times larger than that of OS1 as well.
The periods (126$-$142 min) are roughly equal to that of OS1,
implying that the curvature radius of the magnetic dips supporting the threads is similar.
The damping times are between 170 min and 230 min, and the corresponding quality factor ($\tau/P$) lies in the range of 1.3$-$1.6.
The parameters for OS3 are close to those of longitudinal filament oscillation on 2015 June 29 \citep{zhang17a}.
Different from the longitudinal oscillations (OS2,OS3), the amplitude and the velocity amplitude of the transverse oscillation OS4 are significantly smaller ($\sim$14.7 Mm and 14km s$^{-1}$). OS4 has a period of $\sim$105 min and a damping time of $\sim$142 min, and the corresponding quality factor ($\tau/P$) is $\sim$1.4.
Generally, the damping times of LATOs are in range of 25-180 minutes \citep[tables in][]{Tripathi2009SSRv,shen14a},
while those of LALOs are in range of 115-600 minutes \citep[e.g.,][]{jing03,vrs07}.
Such distinction should be associated with the different damping mechanisms. The damping of LATOs may be attributed to emission of waves or various dissipative processes \citep{Kleczek1969SoPh,Tripathi2009SSRv}. The damping of LALOs are usually caused by the radiation or heat conduction, mass accretion, and wave leakage\citep{zhang12,zhang13,LK12,Luna2012,luna14}.

\subsection{Persistent mass-draining in F1} \label{sec:md}

Figure~\ref{fig3}(b1-b6) shows six snapshots of the EUV 304{~\AA} images during 02:30$-$13:30 UT (see also online movie \textit{OSandDR.mp4}).
The feature traced by the curved line S4 in Figure~\ref{fig3}(b3) indicates that the mass-draining along the barb of F1 toward the chromosphere took place intermittently and lasted for $\sim$14 hr.
In order to investigate the mass-draining,
we used the curved slice S4 (with a length of 280$\arcsec$)
to construct the time-space diagram of Figure~\ref{fig5}(d).
The evolution of mass-draining is roughly divided into four phases.
Before OS1, that is, from $\sim$01:00 UT to $\sim$03:30 UT, the mass-draining is discontinuous with an average velocity of $\sim$35 km s$^{-1}$ (black dotted line).
After the onset of non-damping oscillation OS1 at $\sim$03:35 UT, the mass-draining becomes continuous with a slight increase in velocity (up to $\sim$45 km s$^{-1}$, magenta dotted lines).
It is interesting to note that the amplitude of OS2 and OS4 are larger than the preceding OS1,
at which time the apparent velocity of mass-draining enhances to $\sim$85 km s$^{-1}$ (blue dotted lines) during Eruption3,
which is indicative of filament destabilization.
During 10:00$-$13:00~UT, the mass-draining looks sporadic,
which is due to the fact that the curved slice (S4) does not fully cover the threads guiding the mass-draining in the filament barb.
Finally, the mass-draining reappear with an average velocity of $\sim$85 km s$^{-1}$ (blue dotted lines) during the fast rise and eruption phase of F1.

\begin{figure*}
\centering\includegraphics[width=0.9\textwidth]{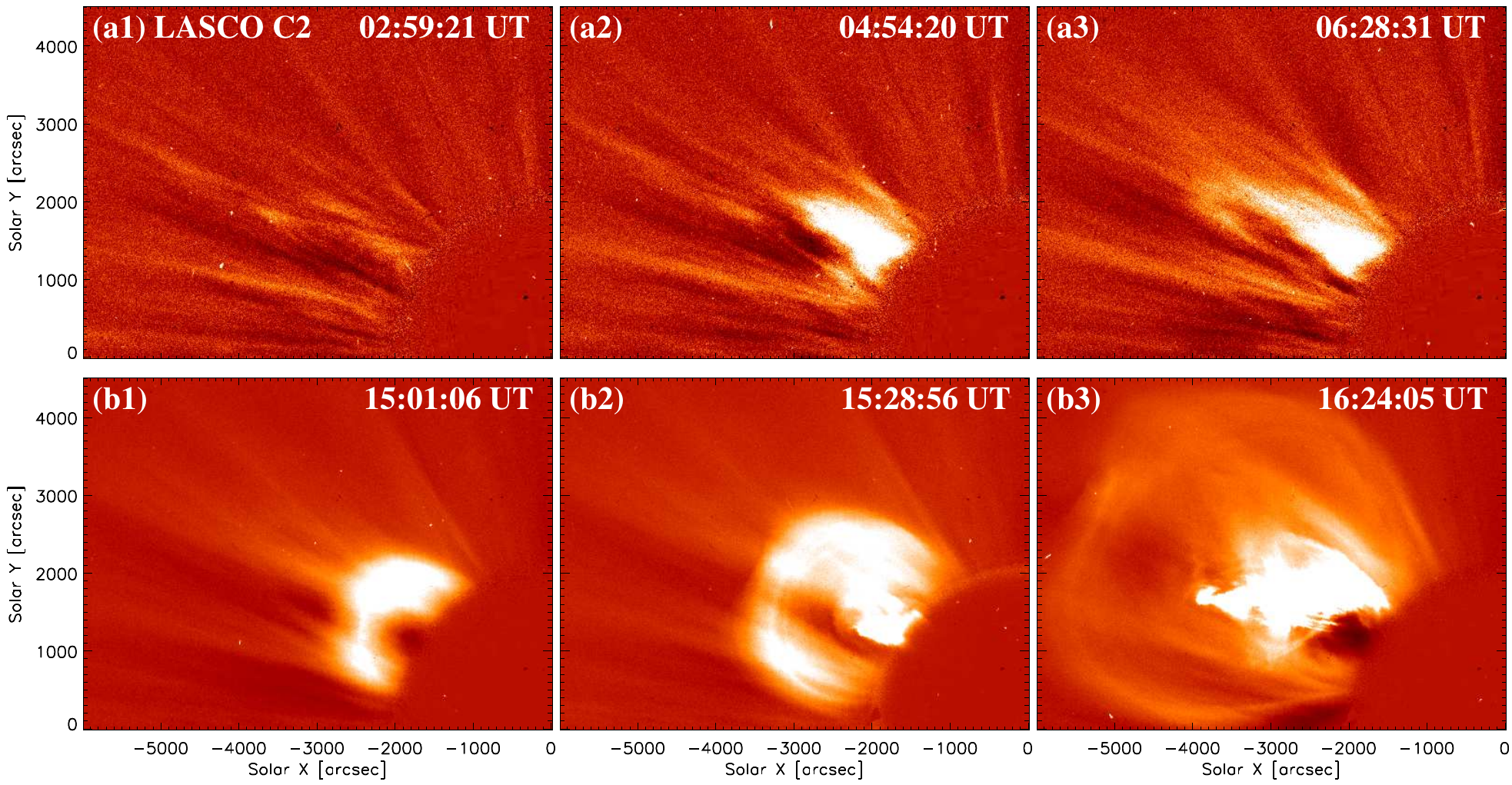}
\caption{Six snapshots of the white-light CME observed by LASCO/C2.
         Top panels show the CME associated with F2 eruption,
         and bottom panels show the CME associated with F1 eruption.
\label{fig7}}
\end{figure*}

\subsection{Eruption of F1} \label{sec:erupt}

Finally, the longer filament F1 erupted successfully in the northeast direction (see Figure~\ref{fig4} and the online movie \textit{F1eruption.mp4}).
The eruption resulted in two flare ribbons as indicated in Figure~\ref{fig4}(d),
leading to a wide CME with a linear speed of $\geq$500 km s$^{-1}$  as observed by LASCO,
in addition to significant coronal dimmings \citep{lor21}.
In Figure~\ref{fig7}(b1-b3),
the CME associated with F1 appeared initially at $\sim$14:29 UT in the FOV of LASCO/C2 and expanded quickly,
showing a typical three-part structure \citep{ill86}.
To investigate the early-phase kinematics of F1,
we select a long straight slice (S5) with a length of 830$\arcsec$ in Figure~\ref{fig4}(c),
and its corresponding time-space diagram is shown in Figure~\ref{fig5}(e).
Likewise, the height of F1 along S5 is outlined with blue diamonds and fitted with Equation~\ref{eqn-1}.
The fitted curve is overlaid with the yellow solid line in Figure~\ref{fig5}(e).
It is seen that F1 started rising slowly from $\sim$02:50 UT with an initial speed of $\sim$1 km s$^{-1}$, and the slow rise lasted for $\sim$500 min.
The onset of the fast rise occurred at $\sim$11:10 UT,
and the average speed of fast rise is $\sim$72 km s$^{-1}$.
In Figure 2(b) of \citet{lor21},
the labeled 126 km s$^{-1}$ represents the final propagation speed of F1 in the FOV of AIA,
which is definitely larger than the average speed,
since the acceleration phase is included in our calculation.

\begin{figure*}
\centering\includegraphics[width=0.9\textwidth]{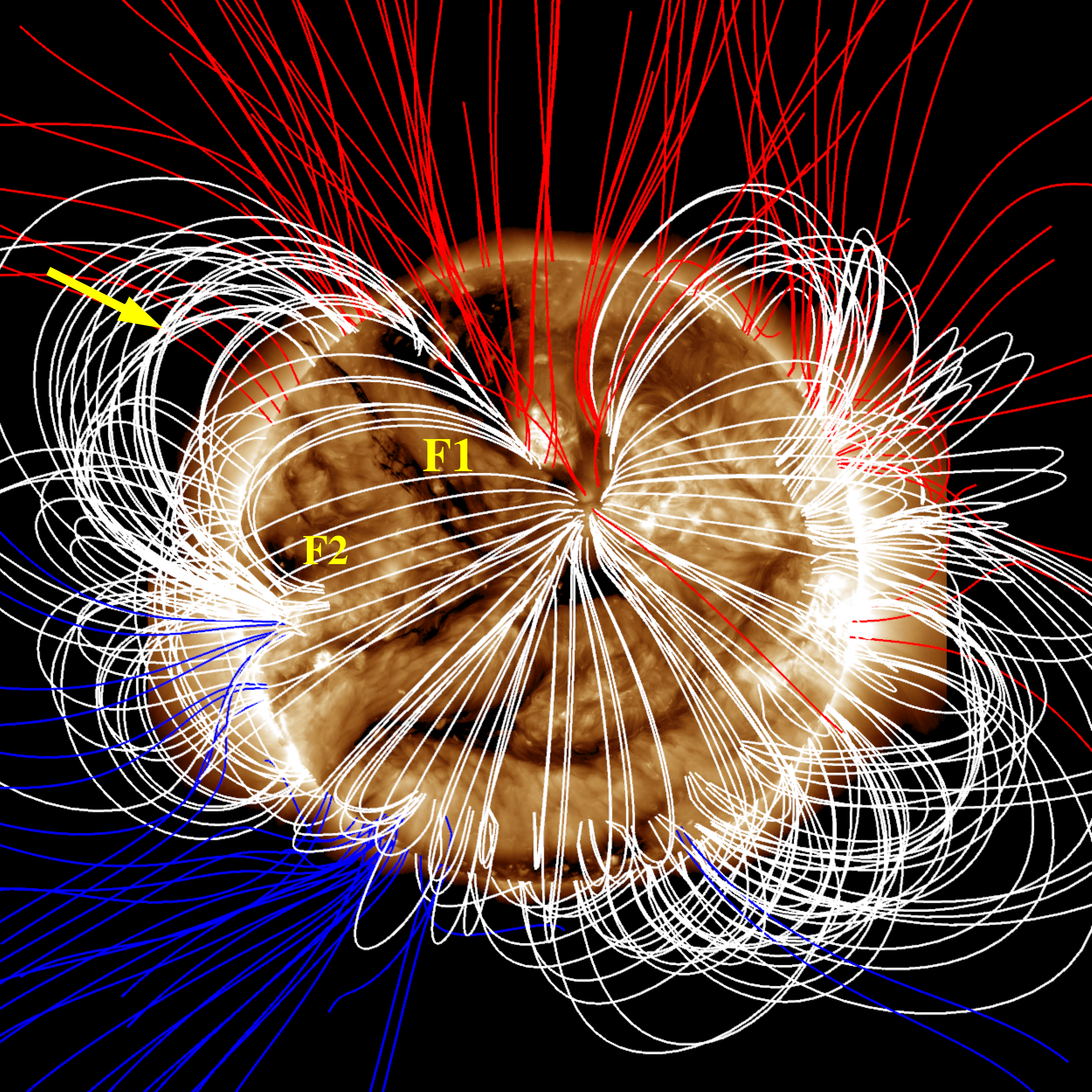}
\caption{The Magnetic field lines obtained by PFSS modeling.
Open and closed field lines are coded with blue/red and white lines, respectively.
The image in AIA 193{~\AA} shows F1 and F2 at 02:00:06 UT.
\label{fig8}}
\end{figure*}
\section{Discussion and Conclusion}\label{sec:dis}

\subsection{Sympathetic eruptions and triggering mechanism of oscillations}

\begin{figure*}
\centering\includegraphics[width=0.9\textwidth]{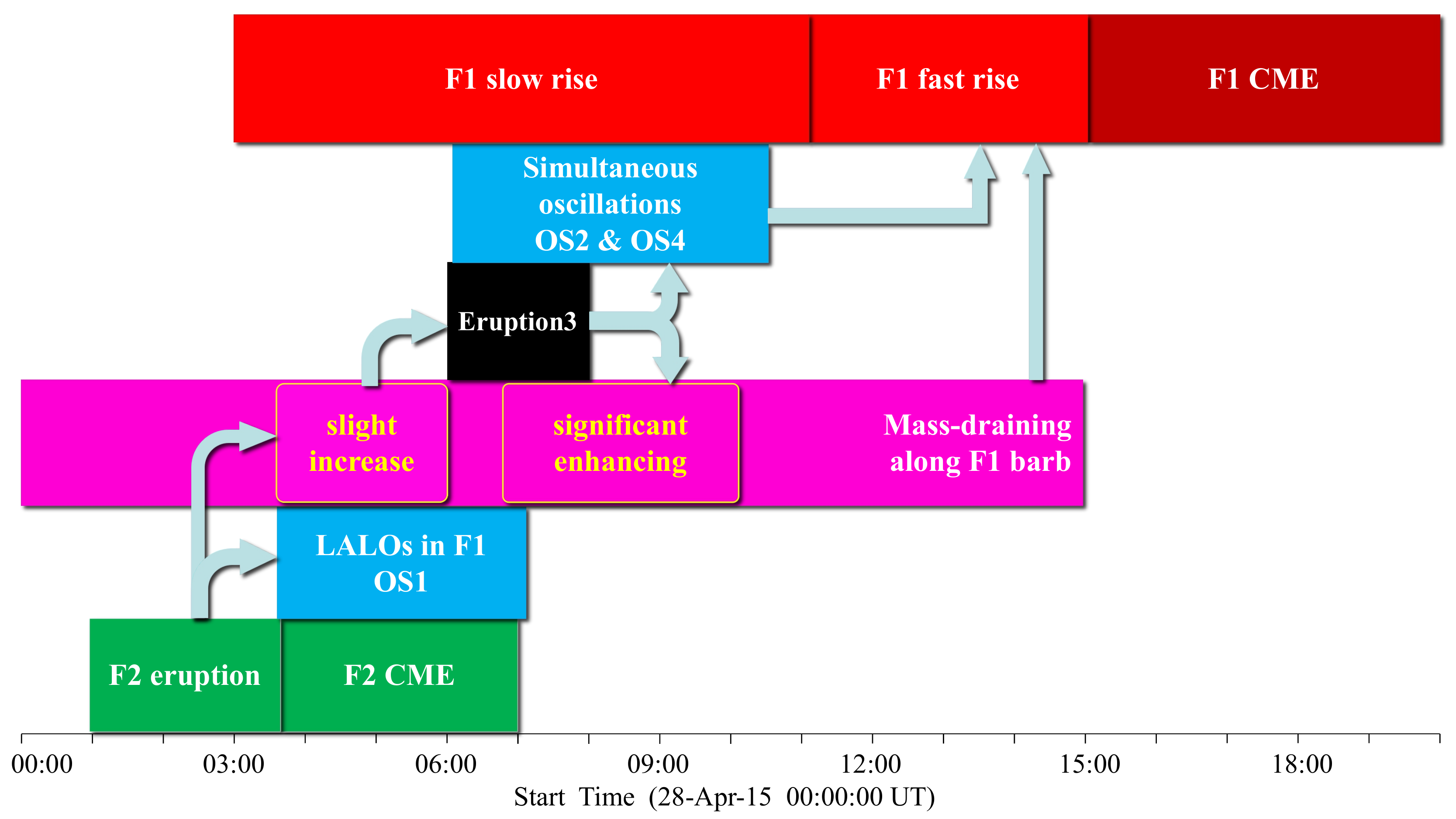}
\caption{Time line of the events on 2015 April 28,
including F2 eruption and the associated CME (green box),
oscillations (blue boxes) and mass-draining (magenta boxes) in F1,
Eruption3 (black boxes), slow rise and fast rise of F1 (red boxes), and CME related to F1 eruption (dark red boxes).
The grey arrows represent cause and effect.
\label{fig9}}
\end{figure*}

In this paper, we report on a sympathetic filament eruption induced by a series of nearby filament eruptions.
The time interval between the first and last eruption is about 8 hours,
longer than those previously reported \citep{shen2012,song2020,hou2020}.

Generally, the main triggering mechanism of LALOs
is sub-flares and micro-flares near the footpoints of filaments
\citep{jing03,jing06,vrs07,LK12,zhang13},
although sometimes nearby filament eruption may excite LALOs \citep{Mazumder2020}.
The green dashed line in Figure~\ref{fig5}(a-c)
denotes the end of the F2 eruption (when it left the FOV at $\sim$03:30 UT) and the start of OS1.
It is seen that the onset of OS1 in F1 is cotemporal with the end time of F2 eruption,
suggesting a connection between OS1 in F1 and F2 eruption.
Furthermore, the 3D magnetic field lines obtained by the PFSS modeling
in Figure~\ref{fig8} show that the two filaments are
almost under the same set of overlying magnetic field lines (marked by the yellow arrow in Figure~\ref{fig8}),
indicating the two filament are closely associated by sharing common overlying magnetic field \citep{cheng2005,tor2011}.
Therefore, the disturbance caused by the F2 eruption would propagate along the magnetic field to F1 \citep{jiang2008},
causing the apparently non-damping LALOs in addition to the slight increase in the mass-draining rate.
Noticeably, F2 eruption caused only longitudinal oscillations but no transverse oscillations,
which may be due to the F2 eruption not producing shock waves \citep{shen14a,shen14b}.

Generally speaking, LATOs could be triggered by the slow-rise or pre-eruption of nearby erupting filaments. \citep{iso06,Isobe2007,pinter2008,chen08}.
The white dashed line in Figure~\ref{fig5}(b-d)
denotes the initiation of Eruption3 and the simultaneous oscillations.
It is obvious that Eruption3 started after the first, slightly enhanced draining,
and Eruption3 is followed closely by both the simultaneous oscillations (OS2 and OS4) and another round of enhanced draining (V3) in F1.
As Eruption3 was projected close to a barb of F1,
the temporal coincidence of the outlined series of events tentatively suggest that Eruption3 may have been involved within the global interplay and occurred in response to the (V2) draining,
before contributing to the triggering of the subsequent (OS2 and OS4) oscillations and additional (V3) draining.
However, without another vantage point from e.g., STEREO \citep{Kaiser2008}, we are unable to elaborate on, or even confirm, this speculative series of events.
Nevertheless, in Figure~\ref{fig9} we draw a time line of the events, showing clearly the
sequence of events potentially associated with the eruption of F1:
\begin{enumerate}
\item{The eruption of F2 is cotemporal with the generation of LALOs and a slight increase of pre-existing mass-draining in F1.
At the same time, F1 started to rise slowly.}

\item{The slight increase of the mass-draining precedes the Eruption3 near the barb of F1.}

\item{Eruption3 is followed by a significant enhancement of the mass-draining and the simultaneous oscillations in F1.}

\item{The persistent draining and oscillations continue up to the successful eruption of the filament F1.}
\end{enumerate}

The complex combination of the aforementioned dynamics may be considered as the collection of processes responsible for bringing the magnetic field of the filament to a point of global instability.
The trigger mechanism of F2 can not be obtained in this study due to insufficient information.

\subsection{Prominence seismology}
Since the first report of LALOs \citep{jing03},
the restoring forces of LALOs have been explored extensively.
Several candidates are proposed, including the magnetic pressure gradient along the filament axis \citep{vrs07, shen14b}, magnetic tension force \citep{LZ12}, gas pressure gradient force \textbf{\citep{jing03},}
and projected gravity along the dip \citep{zhang12}.
Numerical simulations have demonstrated that gravity is the dominant restoring force and
the longitudinal oscillation can be well explained using a pendulum model \citep{LK12,Luna2012,zhang13,fan20}.
For the general pendulum model, the period of oscillation can be expressed as:
\begin{equation} \label{eqn-4}
 P=2\pi \sqrt{\frac{R}{g_\odot}},
\end{equation}
where $R$ is the curvature radius of the dip and $g_\odot=0.274$ km s$^{-2}$ is the gravitational acceleration at photospheric heights.
Using the observed periods of LALOs in Table~\ref{tab-1},
the curvature radius of magnetic dip supporting F1 is estimated to be $\sim$355 Mm,
which is about three times the values established in previous works
\citep{luna2017,zhang17b}.
Besides, the transverse magnetic field strength of the dips can be expressed as \citep{luna14}:
\begin{equation} \label{eqn-6}
B_{tr}[\mathrm{G}]\geq17P[\mathrm{hr}].
\end{equation}
Using the observed value of $P\approx2$ hr, the magnetic field strength at the dips is estimated to be $\geq$34 G,
which is consistent with the value for quiescent prominence derived from the flux rope insertion method \citep[e.g.][]{su12,Guo2021}.

\subsection{Relationship between the longitudinal oscillations, mass-draining, and eruption}
The interplay between mass drainage and longitudinal oscillation in filaments has received relatively little attention.
\citet{bi14} reported on a filament eruption that occurred on 2012 February 23.
Before the eruption, the filament underwent longitudinal oscillations with velocity amplitudes of 54$\pm$3 km s$^{-1}$ and mass drainage toward the solar surface at a speed of $\sim$62 km s$^{-1}$.
The displacement amplitudes of oscillations did not change, but the periods increased by 20\%--30\%,
implying that the dips supporting the filament material became shallower during the oscillations as the result of continuous mass drainage.

In the current study, as indicated in Figure~\ref{fig5},
the mass drainage along the barb started at $\sim$01:00 UT and lasted for $\sim$14 hr,
which is much longer than the durations in the events reported by \citet{bi14} and \citet{zhang17b}.
The apparent velocity of mass unloading increased from $\geq$35 km s$^{-1}$ to $\sim$45 km s$^{-1}$, and later enhanced to $\sim$85 km s$^{-1}$.
The longitudinal oscillations OS1 commenced at $\sim$03:35 UT,
and the simultaneous oscillations (OS2,OS4) started at $\sim$06:05 UT,
with a larger amplitude ($\sim$57 Mm) than OS1.
It suggests there exists a relationship between the sudden and larger-amplitude oscillations and the simultaneous, increased mass-draining,
but we can not ascertain whether the two accelerate each other.
Simultaneous enhancements in mass drainage and oscillations may be manifestations of a gradual destabilization of F1 before its eruption.

Mass-unloading before filament eruption has been noticed in stereoscopic observations \citep{sea11}. \citet{jenk18} concluded that
the depletion of mass from a filament leads to a lighter structure that can then rise to a height that renders the host magnetic field as torus unstable.
The mass depletion reduces the total gravity of the filament and accordingly increases the upward net force that facilitates the final eruption.
In this way, a positive feedback is formed.
In the current study, the density of the filament was $\sim$3.2$\times$10$^{-14}$ g cm$^{-3}$ using the spectroscopic observation \citep{xue2021}, the plane-of-sky velocity of the draining mass was between 35 to 85 km s$^{-1}$, and the cross section of the flux tube in the barb was $\sim$1.65$\times$10$^{18}$ cm$^2$.
Assuming a filling factor of $\sim$0.1 for the filament material along the barb,
the total mass loss during the persistent mass-draining is estimated to have been (1.0--2.2)$\times$10$^{15}$ g, which, depending on the magnitude of the photospheric field beneath the filament, may have played an essential role in triggering the eruption \citep{jenk18,jenk19}.

Using the spectroscopic observation of LATOs within a prominence,
\citet{chen08} proposed that LATOs serves as a precursor of solar eruptions.
\citet{zhang12} proposed that LALOs may also serve as a precursor,
considering that the oscillation may disrupt the equilibrium by changing the plasma distribution.
In current case, both mass-unloading and oscillations are observed before filament eruption.
However, we can not tell which plays a dominant role in triggering the final eruption.
MHD numerical simulations are especially required to distinguish the roles \citep{fan20}.

Due to the lack of stereoscopic observations of the Eruption3,
we are not sure
whether the slight increase of the mass-draining caused Eruption3 and
whether Eruption3 fed back to enhanced the mass-draining.
We only speculate on the possible close connection between them by the simultaneity of these events.
Hence we hope that similar events can be observed form multiple perspectives
in the future.

\subsection{Conclusion}

In this paper, we report a sympathetic successful filament eruption
induced by two nearby eruptions on 2015 April 28.
We give an analysis to the multi-wavelength observations of
the oscillations and mass-draining
within the filament prior to its successful eruption.
Non-damping longitudinal oscillation
and subsequent simultaneous damping oscillations in the larger filament
were cotemporal with the two eruptions, respectively.
The two eruptions are also cotemporal with the enhanced transition of the pre-existing mass-draining along the barb of the filament.
Such cotemporal dynamical relationship is consistent with
the triggering process of the oscillations and enhanced mass-draining
in previous studies.
The events reported here reinforce the theory that
a connection  exists between the nearby eruptions and the oscillations
as well as the mass-draining dynamics.
In light of the contemporaneous nature of the dynamics studied here,
we conclude that their interplay was ultimately responsible for the eruption of filament F1.

\begin{acknowledgements}
SDO is a mission of NASA\rq{}s Living With a Star Program. AIA and HMI data are courtesy of the NASA/SDO science teams.
This work is funded by NSFC grants (No. 11790302, 11790300, 11773079, 41761134088, 11473071) and the Strategic Priority Research Program on Space Science, CAS (XDA15052200, XDA15320301).
Q.M.Z. is also supported by the CAS Key Laboratory of Solar Activity, National Astronomical Observatories (KLSA202006),
and the International Cooperation and Interchange Program (11961131002).
\end{acknowledgements}


\begin{thebibliography}{}
\bibitem[Adrover-Gonz{\'a}lez \& Terradas(2020)]{adro20} Adrover-Gonz{\'a}lez, A. \& Terradas, J.\ 2020, \aap, 633, A113. doi:10.1051/0004-6361/201936841
\bibitem[Arregui et al.(2018)]{arr18} Arregui, I., Oliver, R., \& Ballester, J.~L.\ 2018, Living Reviews in Solar Physics, 15, 3. doi:10.1007/s41116-018-0012-6
\bibitem[Awasthi et al.(2019)]{awa19} Awasthi, A.~K., Liu, R., \& Wang, Y.\ 2019, \apj, 872, 109. doi:10.3847/1538-4357/aafdad
\bibitem[Ballester(2006)]{ball06} Ballester, J.~L.\ 2006, Philosophical Transactions of the Royal Society of London Series A, 364, 405. doi:10.1098/rsta.2005.1706
\bibitem[Bi et al.(2014)]{bi14} Bi, Y., Jiang, Y., Yang, J., et al.\ 2014, \apj, 790, 100. doi:10.1088/0004-637X/790/2/100
\bibitem[Bocchialini et al.(2011)]{bocc11} Bocchialini, K., Baudin, F., Koutchmy, S., et al.\ 2011, \aap, 533, A96. doi:10.1051/0004-6361/201016342
\bibitem[Brueckner et al.(1995)]{bru95} Brueckner, G.~E., Howard, R.~A., Koomen, M.~J., et al.\ 1995, \solphys, 162, 357. doi:10.1007/BF00733434
\bibitem[Chen et al.(2008)]{chen08} Chen, P.~F., Innes, D.~E., \& Solanki, S.~K.\ 2008, \aap, 484, 487. doi:10.1051/0004-6361:200809544
\bibitem[Chen(2011)]{chen11} Chen, P.~F.\ 2011, Living Reviews in Solar Physics, 8, 1. doi:10.12942/lrsp-2011-1
\bibitem[Cheng et al.(2005)]{cheng2005} Cheng, J.-X., Fang, C., Chen, P.-F., et al.\ 2005, \cjaa, 5, 265. doi:10.1088/1009-9271/5/3/006
\bibitem[Cheng et al.(2013)]{cheng13} Cheng, X., Zhang, J., Ding, M.~D., et al.\ 2013, \apjl, 769, L25. doi:10.1088/2041-8205/769/2/L25
\bibitem[Dai et al.(2021)]{dai21} Dai, J., Ji, H., Li, L., et al.\ 2021, \apj, 906, 66. doi:10.3847/1538-4357/abcaf4
\bibitem[Ding et al.(2006)]{ding2006} Ding, J.~Y., Hu, Y.~Q., \& Wang, J.~X.\ 2006, \solphys, 235, 223. doi:10.1007/s11207-006-0092-7
\bibitem[Fan(2020)]{fan20} Fan, Y.\ 2020, \apj, 898, 34. doi:10.3847/1538-4357/ab9d7f
\bibitem[Forbes et al.(2006)]{for06} Forbes, T.~G., Linker, J.~A., Chen, J., et al.\ 2006, \ssr, 123, 251. doi:10.1007/s11214-006-9019-8
\bibitem[Gibson(2018)]{gib18} Gibson, S.~E.\ 2018, Living Reviews in Solar Physics, 15, 7. doi:10.1007/s41116-018-0016-2
\bibitem[Gilbert et al.(2001)]{Gilbert2001ApJ} Gilbert, H.~R., Holzer, T.~E., \& Burkepile, J.~T.\ 2001, \apj, 549, 1221. doi:10.1086/319444
\bibitem[Guo et al.(2021)]{Guo2021} Guo, J., Zhou, Y., Guo, Y., et al.\ 2021, arXiv:2107.12181


\bibitem[Hou et al.(2020)]{hou2020} Hou, Y.~J., Li, T., Song, Z.~P., et al.\ 2020, \aap, 640, A101. doi:10.1051/0004-6361/202038348
\bibitem[Huang et al.(2021)]{huang21} Huang, C.~J., Guo, J.~H., Ni, Y.~W., et al.\ 2021, arXiv:2104.13546
\bibitem[Hyder(1966)]{Hyder1966} Hyder, C.~L.\ 1966, \zap, 63, 78
\bibitem[Illing \& Hundhausen(1986)]{ill86} Illing, R.~M.~E. \& Hundhausen, A.~J.\ 1986, \jgr, 91, 10951. doi:10.1029/JA091iA10p10951
\bibitem[Isobe \& Tripathi(2006)]{iso06} Isobe, H. \& Tripathi, D.\ 2006, \aap, 449, L17. doi:10.1051/0004-6361:20064942
  \bibitem[Isobe et al.(2007)]{Isobe2007} Isobe, H., Tripathi, D., Asai, A., et al.\ 2007, \solphys, 246, 89. doi:10.1007/s11207-007-9091-6
\bibitem[Janvier et al.(2015)]{jan15} Janvier, M., Aulanier, G., \& D{\'e}moulin, P.\ 2015, \solphys, 290, 3425. doi:10.1007/s11207-015-0710-3
\bibitem[Jenkins et al.(2018)]{jenk18} Jenkins, J.~M., Long, D.~M., van Driel-Gesztelyi, L., et al.\ 2018, \solphys, 293, 7. doi:10.1007/s11207-017-1224-y
\bibitem[Jenkins et al.(2019)]{jenk19} Jenkins, J.~M., Hopwood, M., D{\'e}moulin, P., et al.\ 2019, \apj, 873, 49. doi:10.3847/1538-4357/ab037a
\bibitem[Ji et al.(2003)]{ji03} Ji, H., Wang, H., Schmahl, E.~J., et al.\ 2003, \apjl, 595, L135. doi:10.1086/378178
\bibitem[Jiang et al.(2008)]{jiang2008} Jiang, Y., Shen, Y., Yi, B., et al.\ 2008, \apj, 677, 699. doi:10.1086/529417
\bibitem[Jin et al.(2016)]{jin2016} Jin, M., Schrijver, C.~J., Cheung, M.~C.~M., et al.\ 2016, \apj, 820, 16. doi:10.3847/0004-637X/820/1/16
\bibitem[Jing et al.(2003)]{jing03} Jing, J., Lee, J., Spirock, T.~J., et al.\ 2003, \apjl, 584, L103. doi:10.1086/373886
\bibitem[Jing et al.(2006)]{jing06} Jing, J., Lee, J., Spirock, T.~J., et al.\ 2006, \solphys, 236, 97. doi:10.1007/s11207-006-0126-1
\bibitem[Joshi et al.(2016)]{joshi2016} Joshi, N.~C., Schmieder, B., Magara, T., et al.\ 2016, \apj, 820, 126. doi:10.3847/0004-637X/820/2/126
\bibitem[Kaiser et al.(2008)]{Kaiser2008} Kaiser, M.~L., Kucera, T.~A., Davila, J.~M., et al.\ 2008, \ssr, 136, 5. doi:10.1007/s11214-007-9277-0

\bibitem[Kleczek \& Kuperus(1969)]{Kleczek1969SoPh} Kleczek, J. \& Kuperus, M.\ 1969, \solphys, 6, 72. doi:10.1007/BF00146797

\bibitem[Labrosse et al.(2010)]{lab10} Labrosse, N., Heinzel, P., Vial, J.-C., et al.\ 2010, \ssr, 151, 243. doi:10.1007/s11214-010-9630-6
\bibitem[Lemen et al.(2012)]{lemen12} Lemen, J., Title, A., Akin, D., et al.\ 2012, \solphys, 275, 17. doi:10.1007/s11207-011-9776-8
\bibitem[Li \& Zhang(2012)]{LZ12} Li, T. \& Zhang, J.\ 2012, \apjl, 760, L10. doi:10.1088/2041-8205/760/1/L10
\bibitem[Liakh et al.(2020)]{li20} Liakh, V., Luna, M., \& Khomenko, E.\ 2020, \aap, 637, A75. doi:10.1051/0004-6361/201937083
\bibitem[Liakh et al.(2021)]{Liakh2021} Liakh, V., Luna, M., \& Khomenko, E.\ 2021, arXiv:2108.01143

\bibitem[Liu et al.(2009)]{liu2009} Liu, C., Lee, J., Karlick{\'y}, M., et al.\ 2009, \apj, 703, 757. doi:10.1088/0004-637X/703/1/757
\bibitem[Liu et al.(2013)]{liu2013} Liu, R., Liu, C., Xu, Y., et al.\ 2013, \apj, 773, 166. doi:10.1088/0004-637X/773/2/166

\bibitem[Low(1996)]{Low1996SoPh} Low, B.~C.\ 1996, \solphys, 167, 217. doi:10.1007/BF00146338
\bibitem[L{\"o}rin{\v{c}}{\'\i}k et al.(2021)]{lor21} L{\"o}rin{\v{c}}{\'\i}k, J., Dud{\'\i}k, J., Aulanier, G., et al.\ 2021, \apj, 906, 62. doi:10.3847/1538-4357/abc8f6
\bibitem[Luna \& Karpen(2012)]{LK12} Luna, M. \& Karpen, J.\ 2012, \apjl, 750, L1. doi:10.1088/2041-8205/750/1/L1
\bibitem[Luna et al.(2012)]{Luna2012} Luna, M., D{\'\i}az, A.~J., \& Karpen, J.\ 2012, \apj, 757, 98. doi:10.1088/0004-637X/757/1/98
\bibitem[Luna et al.(2014)]{luna14} Luna, M., Knizhnik, K., Muglach, K., et al.\ 2014, \apj, 785, 79. doi:10.1088/0004-637X/785/1/79
\bibitem[Luna et al.(2017)]{luna2017} Luna, M., Su, Y., Schmieder, B., et al.\ 2017, \apj, 850, 143. doi:10.3847/1538-4357/aa9713
\bibitem[Luna et al.(2018)]{luna18} Luna, M., Karpen, J., Ballester, J.~L., et al.\ 2018, \apjs, 236, 35. doi:10.3847/1538-4365/aabde7
\bibitem[Luna \& Moreno-Insertis(2021)]{luna21} Luna, M. \& Moreno-Insertis, F.\ 2021, \apj, 912, 75. doi:10.3847/1538-4357/abec46
\bibitem[Lynch \& Edmondson(2013)]{lynch2013} Lynch, B.~J. \& Edmondson, J.~K.\ 2013, AGU Fall Meeting Abstracts
\bibitem[Mackay et al.(2010)]{mac10} Mackay, D.~H., Karpen, J.~T., Ballester, J.~L., et al.\ 2010, \ssr, 151, 333. doi:10.1007/s11214-010-9628-0
\bibitem[Martin(1998)]{mat98} Martin, S.~F.\ 1998, \solphys, 182, 107. doi:10.1023/A:1005026814076
\bibitem[Mazumder et al.(2020)]{Mazumder2020} Mazumder, R., Pant, V., Luna, M., et al.\ 2020, \aap, 633, A12. doi:10.1051/0004-6361/201936453
\bibitem[Moon et al.(2002)]{moon2002} Moon, Y.-J., Choe, G.~S., Park, Y.~D., et al.\ 2002, \apj, 574, 434. doi:10.1086/340945
\bibitem[Moore et al.(2001)]{Moore2001} Moore, R.~L., Sterling, A.~C., Hudson, H.~S., et al.\ 2001, \apj, 552, 833. doi:10.1086/320559
\bibitem[Moreton \& Ramsey(1960)]{Moreton1960} Moreton, G.~E. \& Ramsey, H.~E.\ 1960, \pasp, 72, 357. doi:10.1086/127549

\bibitem[Okamoto et al.(2004)]{Okamoto2004} Okamoto, T.~J., Nakai, H., Keiyama, A., et al.\ 2004, \apj, 608, 1124. doi:10.1086/420838
\bibitem[Oliver \& Ballester(2002)]{Oliver2002} Oliver, R. \& Ballester, J.~L.\ 2002, \solphys, 206, 45. doi:10.1023/A:1014915428440
\bibitem[Oliver(2009)]{Oliver2009} Oliver, R.\ 2009, \ssr, 149, 175. doi:10.1007/s11214-009-9527-4
\bibitem[Pant et al.(2015)]{Pant2015RAA} Pant, V., Srivastava, A.~K., Banerjee, D., et al.\ 2015, Research in Astronomy and Astrophysics, 15, 1713. doi:10.1088/1674-4527/15/10/008
\bibitem[Parenti(2014)]{par14} Parenti, S.\ 2014, Living Reviews in Solar Physics, 11, 1. doi:10.12942/lrsp-2014-1
\bibitem[Pint{\'e}r et al.(2008)]{pinter2008} Pint{\'e}r, B., Jain, R., Tripathi, D., et al.\ 2008, \apj, 680, 1560. doi:10.1086/588273
\bibitem[Pesnell et al.(2012)]{pesn12} Pesnell, W., Thompson, B., \& Chamberlin, P.\ 2012, \solphys, 275, 3
\bibitem[R{\'e}gnier et al.(2011)]{reg11} R{\'e}gnier, S., Walsh, R.~W., \& Alexander, C.~E.\ 2011, \aap, 533, L1. doi:10.1051/0004-6361/201117381
\bibitem[Ramsey \& Smith(1966)]{Ramsey1966} Ramsey, H.~E. \& Smith, S.~F.\ 1966, \aj, 71, 197. doi:10.1086/109903
\bibitem[Ruderman \& Luna(2016)]{rud16} Ruderman, M.~S. \& Luna, M.\ 2016, \aap, 591, A131. doi:10.1051/0004-6361/201628713

\bibitem[Schou et al.(2012)]{schou12} Schou, J., Scherrer, P., Bush, R., et al.\ 2012, \solphys, 275, 229. doi:10.1007/s11207-011-9842-2

\bibitem[Schrijver \& Title(2011)]{Schrijver2011} Schrijver, C.~J. \& Title, A.~M.\ 2011, Journal of Geophysical Research (Space Physics), 116, A04108. doi:10.1029/2010JA016224
\bibitem[Schrijver et al.(2013)]{Schrijver2013} Schrijver, C.~J., Title, A.~M., Yeates, A.~R., et al.\ 2013, \apj, 773, 93. doi:10.1088/0004-637X/773/2/93
\bibitem[Schrijver \& De Rosa(2003)]{Schrijver2003SoPh} Schrijver, C.~J. \& De Rosa, M.~L.\ 2003, \solphys, 212, 165. doi:10.1023/A:1022908504100


\bibitem[Seaton et al.(2011)]{sea11} Seaton, D.~B., Mierla, M., Berghmans, D., et al.\ 2011, \apjl, 727, L10. doi:10.1088/2041-8205/727/1/L10
\bibitem[Shen et al.(2012)]{shen2012} Shen, Y., Liu, Y., \& Su, J.\ 2012, \apj, 750, 12. doi:10.1088/0004-637X/750/1/12
\bibitem[Shen et al.(2014a)]{shen14a} Shen, Y., Ichimoto, K., Ishii, T.~T., et al.\ 2014, \apj, 786, 151. doi:10.1088/0004-637X/786/2/151
\bibitem[Shen et al.(2014b)]{shen14b} Shen, Y., Liu, Y.~D., Chen, P.~F., et al.\ 2014, \apj, 795, 130. doi:10.1088/0004-637X/795/2/130
\bibitem[Shen et al.(2017)]{shen17} Shen, Y., Liu, Y., Tian, Z., et al.\ 2017, \apj, 851, 101. doi:10.3847/1538-4357/aa9af0
\bibitem[Song et al.(2020)]{song2020} Song, Z., Hou, Y., Zhang, J., et al.\ 2020, \apj, 892, 79. doi:10.3847/1538-4357/ab77b3
\bibitem[Su \& van Ballegooijen(2012)]{su12} Su, Y. \& van Ballegooijen, A.\ 2012, \apj, 757, 168. doi:10.1088/0004-637X/757/2/168

\bibitem[Terradas et al.(2015)]{ter15} Terradas, J., Soler, R., Luna, M., et al.\ 2015, \apj, 799, 94. doi:10.1088/0004-637X/799/1/94
\bibitem[Titov et al.(2012)]{Titov2012} Titov, V.~S., Mikic, Z., T{\"o}r{\"o}k, T., et al.\ 2012, \apj, 759, 70. doi:10.1088/0004-637X/759/1/70
\bibitem[T{\"o}r{\"o}k \& Kliem(2005)]{tor2005} T{\"o}r{\"o}k, T. \& Kliem, B.\ 2005, \apjl, 630, L97. doi:10.1086/462412

\bibitem[T{\"o}r{\"o}k et al.(2011)]{tor2011} T{\"o}r{\"o}k, T., Panasenco, O., Titov, V.~S., et al.\ 2011, \apjl, 739, L63. doi:10.1088/2041-8205/739/2/L63
\bibitem[Tripathi et al.(2009)]{Tripathi2009SSRv} Tripathi, D., Isobe, H., \& Jain, R.\ 2009, \ssr, 149, 283. doi:10.1007/s11214-009-9583-9
\bibitem[van Ballegooijen \& Martens(1989)]{vanb89} van Ballegooijen, A.~A. \& Martens, P.~C.~H.\ 1989, \apj, 343, 971. doi:10.1086/167766
\bibitem[Vr{\v{s}}nak et al.(2007)]{vrs07} Vr{\v{s}}nak, B., Veronig, A.~M., Thalmann, J.~K., et al.\ 2007, \aap, 471, 295. doi:10.1051/0004-6361:20077668
\bibitem[Wang et al.(2001)]{wang2001} Wang, H., Chae, J., Yurchyshyn, V., et al.\ 2001, AGU Spring Meeting Abstracts
\bibitem[Wang et al.(2016)]{wang2016} Wang, R., Liu, Y.~D., Hu, H., et al.\ 2016, AGU Fall Meeting Abstracts
\bibitem[Wang et al.(2018)]{wang2018} Wang, D., Liu, R., Wang, Y., et al.\ 2018, \apj, 869, 177. doi:10.3847/1538-4357/aaef35
\bibitem[Wheatland \& Craig(2006)]{wheatland2006} Wheatland, M.~S. \& Craig, I.~J.~D.\ 2006, \solphys, 238, 73. doi:10.1007/s11207-006-0206-2
    \bibitem[Xue et al.(2021)]{xue2021} Xue, J., Li, H., \& Su, Y.\ 2021, arXiv:2109.02908
\bibitem[Zhang et al.(2012)]{zhang12} Zhang, Q.~M., Chen, P.~F., Xia, C., et al.\ 2012, \aap, 542, A52. doi:10.1051/0004-6361/201218786
\bibitem[Zhang et al.(2013)]{zhang13} Zhang, Q.~M., Chen, P.~F., Xia, C., et al.\ 2013, \aap, 554, A124. doi:10.1051/0004-6361/201220705
\bibitem[Zhang et al.(2017a)]{zhang17a} Zhang, Q.~M., Li, D., \& Ning, Z.~J.\ 2017a, \apj, 851, 47. doi:10.3847/1538-4357/aa9898
\bibitem[Zhang et al.(2017b)]{zhang17b} Zhang, Q.~M., Li, T., Zheng, R.~S., et al.\ 2017b, \apj, 842, 27. doi:10.3847/1538-4357/aa73d2
\bibitem[Zhang \& Ji(2018)]{zhang2018} Zhang, Q.~M. \& Ji, H.~S.\ 2018, \apj, 860, 113. doi:10.3847/1538-4357/aac37e
\bibitem[Zhang et al.(2019)]{zhang19} Zhang, L.~Y., Fang, C., \& Chen, P.~F.\ 2019, \apj, 884, 74. doi:10.3847/1538-4357/ab3d3a
\bibitem[Zhang et al.(2020)]{zhang20} Zhang, Q.~M., Guo, J.~H., Tam, K.~V., et al.\ 2020, \aap, 635, A132. doi:10.1051/0004-6361/201937291
    \bibitem[Zhang et al.(2021)]{zhang2021arXiv} Zhang, Q., Liu, R., Wang, Y., et al.\ 2021, arXiv:2108.09401
\bibitem[Zhou et al.(2017)]{zhou17} Zhou, Y.-H., Zhang, L.-Y., Ouyang, Y., et al.\ 2017, \apj, 839, 9. doi:10.3847/1538-4357/aa67de
\bibitem[Zhou et al.(2018)]{zhou18} Zhou, Y.-H., Xia, C., Keppens, R., et al.\ 2018, \apj, 856, 179. doi:10.3847/1538-4357/aab614
\bibitem[Zhukov \& Veselovsky(2007)]{zhukov2007} Zhukov, A.~N. \& Veselovsky, I.~S.\ 2007, \apjl, 664, L131. doi:10.1086/520928

\end{thebibliography}
\end{document}